\documentclass{article}

\usepackage{amssymb,amsfonts,amsmath,stmaryrd}
\usepackage{cite,enumerate,float,indentfirst}
\usepackage{color}
\usepackage{tikz}
\usetikzlibrary{arrows,snakes,backgrounds}
\usepackage{url}
\usepackage{hyperref}
\usepackage{textcomp}

\def\be{\begin{eqnarray}}
\def\ee{\end{eqnarray}}
\def\nn{\nonumber}

\def\tr{{\rm tr}\,}
\def\Tr{{\rm Tr}\,}

\definecolor{red}{rgb}{1,0,0}
\definecolor{orange}{rgb}{1,0.5,0}
\definecolor{violet}{rgb}{0.7,0,1}

\hoffset= - 1.0in         

\begin{document}

\title{\vspace{1cm}{\Large {\bf
      Superintegrability of matrix Student's distribution
    }
    \date{}
    \author{
    {\bf A. Mironov$^{a,b,c,\ddag}$}
      {\bf A. Morozov$^{b,c,d,*}$}
      {\bf A. Popolitov$^{b,c,d,\dag}$}}
  }}

\maketitle
\vspace{-5.5cm}

\begin{center}
\hfill FIAN/TH-11/21 \\
	\hfill ITEP/TH-19/21 \\
	\hfill IITP/TH-14/21 \\
	\hfill MIPT/TH-13/21
\end{center}

\vspace{3.cm}

\begin{center}
$^a$ {\small {\it Lebedev Physics Institute, Moscow 119991, Russia }} \\
  $^b$ {\small {\it Institute for Theoretical and Experimental Physics, Moscow 117218, Russia}}\\
  $^c$ {\small {\it Institute for Information Transmission Problems, Moscow 127994, Russia}}\\
  $^d$ {\small {\it Moscow Institute of Physics and Technology, Dolgoprudny 141701, Russia }} \\
  \vspace{0.25cm}
 $\ddag$\;{\small {\it mironov@itep.ru; mironov@lpi.ru}}\ \ \ \ \ \  $*$\;{\small {\it morozov.itep@mail.ru}}
  \ \ \ \ \ \ $\dag$\;{\small {\it popolit@gmail.com}}
\end{center}

\vspace{.5cm}

\begin{abstract}
For ordinary matrix models, the eigenvalue probability density decays rapidly
  as one goes to infinity, in other words, has ``short tails''.
  This ensures that all the multiple trace correlators (multipoint moments)
  are convergent and well-defined.
  Still, many critical phenomena are associated with an enhanced probability of
  seemingly rare effects,
  and one expects that they are better described by the "long tail" models.
  In absence of the exponential fall-off, the integrals for high moments diverge,
  and this could  imply a loss of (super)integrability properties
  pertinent  to matrix and eigenvalue models and, presumably, to  non-perturbative
  (exact) treatment of more general quantum systems.
  In this paper, we explain that this danger to modern understanding could be exaggerated.
  We consider a simple  family of long-tail matrix models,
  which preserve the crucial feature of superintegrability:
  exact factorized expressions for a full set of basic averages.
  It turns out that superintegrability can survive
  after an appropriate (natural and obvious) analytical continuation
  even in the presence of divergencies,
  which opens new perspectives for the study of the long-tail matrix models.
\end{abstract}

\bigskip

\bigskip

{\section{Introduction} \label{sec:introduction}

  In abstract theory, the thermodynamical equilibrium is associated with the
  Gaussian distribution \cite{book:LL-statistical-physics}.
  Still, it is well-known that, in practice, many relevant random distributions
  do not behave this way:
  strong deviations from the average appear much more often than expected,
  and are suppressed by powers rather than exponentially.
  In other words, the true distributions often have ``long tails''.
  This is well-known about earthquakes \cite{paper:SCP-tsallis-entropy-earthquakes,
    paper:CPLRV-self-organized-criticality-and-earthquakes},
  stock trade \cite{paper:TD-a-map-and-simple-heuristic-to-detect-fragility-antifragility-and-model-error},
  flicker (pink) noise \cite{paper:MS-on-1f-noise-and-other-distributions-with-long-tails},
  self-organized criticality \cite{bak1987self,book:J-self-organized-criticality}
  and many other examples, which can inspire a creation of entire new paradigms
  \cite{haken1981erfolgsgeheimnisse,book:T-antifragile}.

  In fact, this is well-known even in the undergraduate course
  of experimental physics:
  the true distributions of random data obey the long-tail Student distribution
  rather than the classical Gaussian one.
  Also well-known is an explanation:
  since most results tend to lie around the peak,
  one can easily {\it under}estimate the true dispersion,
  and get an illusion that it is small:
  then with more measurements one observes stronger deviations from the putative mean
  than originally anticipated, and they seem \textit{not} to be exponentially damped.
  The reason is that the true dispersion is bigger,
  but one needs (exponentially) more data to see this.
 In formal language, the reason is that one actually measures
  not the true Gaussian distribution, but that of the ratio of
  Gaussian distributed quantities, which is characterized
  by another, {\it Student } distribution \cite{paper:S-the-probable-error-of-a-mean}.
  It is power-like for any number $n$ of measurements,
  and tends to the Gaussian distribution only when $n\longrightarrow \infty$.
  The Student distribution is therefore a nice representative from the
  long-tail class.
  Still, it is deduced from the Gaussian one, and one can expect
  that the more sophisticated (or more abstract) properties will survive.

  In this letter, we explain that this expectation is not far from truth:
  the Student  distribution remains {\it superintegrable}, i.e. all the
  correlators are given by nicely factorized rational functions.

  At the same time, these functions can be negative!
  Of course, this is impossible for {\it physically meaningful} quantities:
  once the underlying Gaussian hidden variables are restored, both the poles
  and the negativity disappear.
  Nevertheless, for the {\it formal} distribution this does happen:
  integrals in non-physical domain formally start to diverge,
  and an analytical continuation preserves the factorization at the expense of positivity,
  i.e. reveals some other layer of physical reality and requires a new interpretation.
  Most importantly, however, we demonstrate that there is no contradiction between
  long tails and the modern non-perturbative quantum theory,
  where the invariance of exact functional integrals under arbitrary change
  of integration variables (fluctuating fields) reflects itself in the Ward identities
 often giving rise to integrability properties of effective actions.

 We will mainly consider matrix Student distributions in order to have more degrees of freedom so that the theory
 would be closer to real complicated systems. This kind of long-tail distributions was also studied previously in various
 applications, from financial markets to very theoretical string issues, see e.g. \cite{paper:TUMMINELLO201040,AKEMANN20102566,Bouchaud,Johnstone,CMS,Stanford:2019vob}.
}

{\section{Student distribution} \label{sec:student-distribution}

  The density of the Student  distribution
  \cite{
    paper:S-the-probable-error-of-a-mean
  } is given by the following formula
  \begin{align} \label{eq:one-dim-student}
   \boxed{
    d \mu(z) =  \frac{dz}{\left (1 + \frac{z^2}{a^2}\right)^\alpha}
  }
  \end{align}
  and it is apparently ``long-tail" as compared to the Gaussian distribution  $e^{-x^2}dx$.
Nevertheless, it is nothing but the distribution of the ratio of two Gaussian variables, $X$ and $Y$:
$$
\frac{d\mu^{(1)}(z)}{dz} = \int \delta\left(t - \frac{X}{Y}\right)e^{-X^2-Y^2} dX dY
\sim \int e^{i\alpha\left(z- \frac{X}{Y}\right)}e^{-X^2-Y^2}d\alpha\, dX dY
= \int e^{i\alpha z} e^{-\alpha^2/4Y^2}e^{ -Y^2} d\alpha \,dY
\sim
$$
\be
\sim 2\int |Y| e^{-(1+z^2)Y^2} dY = \frac{1}{1+z^2}
\label{st1}
\ee
and the long tail arises because the denominator can be much smaller than the numerator,
  while the both remain near the Gaussian peaks, and are not exponentially damped.
Indeed, $z$ exceeds some large value $z_0$ whenever $Y<1/z_0$, and the integral
$\int_0^{1/z_0} dY e^{-Y^2} \sim \frac{1}{z_0} = \int_{z_0}^\infty d\mu^{(1)}(z)$.

More general is the situation, when $\vec Y$ is an $n$-component vector,
a collection of $n$ Gaussian variables,
while $Y:=\sqrt{\frac{1}{n}\sum_{i=1}^{n} Y_i^2}$.
Then the degree of the power-like distribution changes,
but the long-tail feature persists:
\begin{align} \label{eq:vector-student}
\frac{d\mu^{(n)}(z)}{dz} = \int \delta\left(z - \frac{X}{Y}\right)e^{-X^2-\vec Y^2} dX d^nY
\sim & \int e^{i\alpha\left(z- \frac{Y_0}{Y}\right)}e^{-X^2-\vec Y^2}d\alpha\, dX d^nY
= \int e^{i\alpha z} e^{-\alpha^2/4Y^2} e^{ -\vec Y^2} d\alpha\, d^nY
\sim
\\ \notag
\sim & \int |Y| e^{-\left(1+\frac{z^2}{n}\right)\vec Y^2} d^nY
\sim \frac{1}{\left(1+\frac{z^2}{n}\right)^{\frac{n+1}{2}}}
\end{align}

The Student's distribution plays a great role in experimental sciences
  because it naturally arises when one tries to estimate the mean value of a Gaussian random variable
  $x \sim \mathcal{N}(\nu,\sigma)$
  from the sample of its measurements $x_0, \dots, x_n$,
  but does not know the dispersion $\sigma$ in advance.
  Then, since most results are near the Gaussian peak, it is easy to underestimate the true dispersion,
  before the truly rare events show up and demonstrate that the dispersion is bigger than it
  originally seemed.

\smallskip

\noindent At a more formal level, let
\be
  \overline{x} = & \frac{1}{n+1} \sum_{i=0}^n x_i \nn\\
  \nn\\
  S^2 = & \frac{1}{n} \left( \sum_{i=0}^n x_i^2   - (n+1) \overline{x}^2 \right )
\ee
be the average and the mean quadratic deviation defined from the sample.
The question is how the variable $z:=\frac{\sqrt{n} \cdot \overline{x}}{S}$ is distributed.
For simplicity, we put the mean value $\nu$ of the Gaussian distributed variable $x$ to be zero.
Then the relevant density
$d \mu(z)$ is equal to
\be \label{eq:multidim-student}
 \frac{ d\mu(z)}{dz} =  \int d x_0 \dots d x_n \exp \left ( -\frac{1}{2 \sigma} \sum_{i=0}^n x_i^2 \right )
  \cdot \delta \left ( z - \frac{\sqrt{n} \cdot \overline{x}}{S} \right )
\ee
One can perform an orthogonal change of integration variables such that the first integration variable becomes
\be
 X := \frac{1}{\sqrt{n+1}} \sum_{i=0}^n x_i
\ee
Denoting as $\vec Y$ the set of $x$-variables orthogonal to this $X\sim \bar x$,
one returns to the calculation (\ref{eq:vector-student}).
Note that $\sigma$ drops out from the final formula.

Since the Student distribution does not decay fast enough at infinity:
$d \mu(z) \sim O(|z|^{-2\alpha})$ as $|z| \rightarrow \infty$,
its moments $\left\langle z^n \right \rangle = \int \mu(z) z^n$ do not converge starting from certain $n$.
Therefore one needs \textit{extra} considerations to regularize/define the divergent integrals.
One of the ways to do this is through some matrix generalization of \eqref{eq:one-dim-student}, which
we introduce in the next Section. We discuss various subtleties and implications of this approach in Section~\ref{sec:caveats}.
}

{\section{Matrix Student distribution} \label{sec:matrix-student-distribution}

  The matrix generalization of \eqref{eq:one-dim-student} is straightforward.
  It even appeared in the literature
  \cite{book:GN-matrix-variate-distributions},
  but with no relation to integrability
  and other hidden structures typical for matrix models.
  Let $X$ and $Y$ be rectangular real-valued matrices
  of sizes $N \times M$  and   $N \times (N+n - 1)$ respectively.
Consider the following density function for their peculiar combination
$Z = \left ((Y Y^T)^{1/2}\right)^{-1} X$, which is a rectangular $N\times M$ matrix:
\begin{align} \label{eq:mat-student-density}
\frac{d\mu(Z)}{  dZ}= \int dX dY \delta \left ( Z -  (Y Y^T)^{-1/2}  X \right )
  \exp \left ( -\frac{1}{2} \tr A^{-1} Y Y^T -\frac{1}{2} \tr X B X^T\right ),
\end{align}
where $A$ and $B$ are constant square $N\times N$ and $M\times M$ matrices parameterizing  the model.
The inverse square root is taken
in the sense of quadratic form, not of linear operator:
\begin{align}
  (Y Y^T)^{1/2} \left((Y Y^T)^{1/2}\right)^T = Y Y^T
\end{align}

\noindent Performing integration in $X$, using the $\delta$-function, and taking into account the relevant Jacobian,
one gets
\begin{align} \label{eq:mat-student-density-continued}
  \eqref{eq:mat-student-density} &
  \mathop{=}_{\int_X}   \int dY \exp \tr \left ( -\frac{1}{2} Z B Z^T Y Y^T - \frac{1}{2} A^{-1} Y Y^T \right )
  \det \left ( Y Y^T \right )^{\frac{M}{2}}
\end{align}

Now, rewriting the determinant as an integral over auxiliary
$M/2$ copies of complex Grassmann variables
\begin{align}
  \det \left ( Y Y^T \right )^{\frac{M}{2} } \sim \int d \theta d \bar\theta
  \exp \tr \left ( - \sum_{i=1}^N \sum_{j=1}^{M/2}\bar\theta_{i,j} Y Y^T \theta_{i,j} \right )
\end{align}
one can perform integral over $Y$
\begin{align}
  \eqref{eq:mat-student-density-continued} & \mathop{=}
  \int   d\theta d\bar\theta
  \det \left ( Z B Z^T + A^{-1} + \theta \theta^\dagger \right )^{-\frac{N +n  - 1}{2} } =
  \\ \notag = &
    \det \left ( Z B Z^T + A^{-1} \right )^{-\frac{1}{2}(n+M-1)}
  \int d\theta d\bar\theta \det \left ( I_N + \left(Z B Z^T + A^{-1}\right)^{-1} \theta \theta^\dagger \right )^{-\frac{N+n-1}{2}} =
  \\ \notag \mathop{=}  &
     \det \left ( Z B Z^T + A^{-1} \right )^{-\frac{1}{2}(N+M+n-1)}
  \sim   \det \left ( I_N + A Z B Z^T \right )^{-\frac{1}{2}(N+M+n-1)},
\end{align}
where the last Grassmann integral is taken by using diagonalization.
The result is the desired matrix generalization of the Student distribution.

In what follows, we  concentrate on the simplest form of this matrix   Student  distribution,
with $M=N$, when  $Z$ is a square matrix,  $B = I$, $A = \frac{1}{a^2} I$
and $n = 2 \alpha + 1 - 2 N$.
Moreover, we take $Z$ to be Hermitian matrix instead of an arbitrary real-valued matrix,
the derivation of the integration measure being similar.
The case of rectangular non-Hermitian $Z$, and of arbitrary matrices $A$ and $B$,
will be considered elsewhere.

For the unit $A$ and $B$ matrices, it is obvious that the Student matrix model
\be\label{partf0}
\boxed{
\int_{N\times N} \frac{dZ}{\det \left ( I +  Z ^2\right )^{N+\frac{n-1}{2}}}
}
\ee
depends only on the eigenvalues of the $N\times N$ square matrix-variable $Z$,
and it possesses the usual properties of matrix models
\cite{
  paper:M-matrix-models-as-integrable-systems,
  paper:M-integrability-and-matrix-models,
  paper:M-2dgravity,
  paper:M-quantum-deformations-of-tau-functions
},
of which we discuss just two: Ward identities \cite{paper:David,paper:MM-Virasoro,paper:AM-Virasoro,paper:IM-Virasoro}
and superintegrability \cite{
  paper:MM-on-the-complete-perturbative-solution-of-on-matrix-models,
  paper:MM-correlators-in-tensor-models-from-character-calculus,  paper:MM-sum-rules-for-characters-from-characters-preservation-property-of-matrix-models}. The third basic property, integrability \cite{paper:GMMMO,paper:KMMOZ} is trivially presented since
it is a formal property of the one-matrix integral over the Hermitian matrix with an invariant measure, and the matrix model (\ref{partf0}) is exactly of this type.

These properties correspond to introduction of sources/deformations in the partition function (\ref{partf0})
but in two slightly different ways.
Also, as we explain in sec.6, though the results are formally valid for all values of $\alpha$ and $N$,
they should be treated with care to be physically sensible.

{\section{Ward identities} \label{sec:ward-identities}

In terms of eigenvalues, the measure of the simplest matrix Student distribution reads
  \begin{align} \label{eq:matrix-student-eigenvalues}
      d \mu = dz_1 \dots dz_N \cdot \prod_{i<j}^N\left(z_i-z_j\right)^2 \cdot
      \prod_{i=1}^N \frac{1}{\det\left(1 + \frac{z_i^2}{a^2} \right)^\alpha}
  \end{align}
  and the averages and the power-sum (multi-trace) correlators are defined as usual
  \begin{align}
    C_{i_1, \dots, i_m} :=
    \Big\langle \, \Tr Z^{i_1}\ldots \Tr Z^{i_m}\Big\rangle \  =  \
    \left\langle\sum_{j_1}z_{j_1}^{i_1} \dots \sum_{j_k}z_{j_m}^{i_m} \right\rangle
  \end{align}
with the average $\left\langle f(Z) \right\rangle := \frac{\int f(Z) d\mu}{\int d\mu}$.
Inserting a suitably chosen set of full-derivatives in the integrand
  \begin{align}
\sum_{j=1}^N \int  \frac{\partial}{\partial z_j}\left\{ \left(1 + \frac{z_j^2}{a^2}\right) z_j^n
   \ \Tr Z^{i_1}\ldots \Tr Z^{i_m}\  d\mu
   \right\}  = 0
  \end{align}
  one  obtains a set of Ward identities for the correlators
  \begin{align} \label{eq:ward-identities}
    &
    \boxed{
      \sum_{k = 0}^{n-1} C_{k, n-1-k, I}
      + \frac{1}{a^2} \sum_{k = 0}^{n+1} C_{k, n+1-k, I}
      - \frac{2 \alpha}{a^2} C_{n+1, I}
      + \sum_{s=1}^m \left (
      i_s \,      C_{ i_s + n - 1,\,I \backslash i_s}
      + \frac{i_s}{a^2}\, C_{ i_s + n + 1,\, I \backslash i_s}
      \right ) = 0
    }
  \end{align}
  where  $I$  denotes a multi-index $i_1, \dots i_m$, and
  $\backslash i_s$ means deletion of the element $i_s$ from the multi-index.
  All integrals are defined in the sense of the principal value,
  which is important in order to eliminate possible contributions of the boundary terms
  at infinities.

  Actually, these constraints are as strong as the ones for the Gaussian Hermitian model:
  if one starts from initial
  conditions $C_\emptyset = 1$, \ \ $C_{1} = 0$, one can unambiguously find
  every correlator $C_{i_1,\dots,i_m}$ in a finite amount of steps.
Specifically, to find every correlator with a sum over indices, say, $p$
  provided all the correlators up to (including) degree $p-2$ are already known,
  one needs to consider equations \eqref{eq:ward-identities} with
  \begin{align}
    n + |I| = p - 1
  \end{align}
  For instance, in order to find the simplest non-trivial correlators $C_2$ and $C_{1,1}$,
  one needs to consider two equations: those with $n=1$ and $I = \emptyset$, and with $n=0$
  and $I = [1]$.
}
}

{\section{Superintegrability} \label{sec:superintegrability}

The fact that it is possible to define seemingly-divergent matrix-model correlators
of the long-tail Student model
in an integrable way, i.e. in such a way that they are consistent with
an infinite system of Ward identities, which determines them unambiguously
is already surprising enough, and has interesting implications for predictability of risks, avalanches and earthquakes.
However,  there is even more: correlators of the model can be found
once and for all, in other words, the model is \textit{superintegrable}
\cite{
  paper:MM-sum-rules-for-characters-from-characters-preservation-property-of-matrix-models,
  paper:IMM-tensorial-generalization-of-characters
}.
As usual, this requires a switch from $C_I$ to an appropriate basis in the space of
correlation functions.
Namely, the correlators of Schur polynomials $\chi_\lambda$,
which are peculiar linear combinations of the
power-sum (multi-trace) correlators enumerated by the Young-diagram $\lambda$,
are simple factorized expressions, again in terms of the Schur polynomials
with just the same $\lambda$,
depending now on the matrix size $N$. The first examples are:
  \be
    \left \langle \chi_{[2]} \right \rangle =
    \frac{a^2 N (N+1)}{2 (2 (\alpha - N) - 1)}
& \ \ \ \ \ \ \
&    \left \langle \chi_{[1,1]} \right \rangle =
    -\frac{a^2 (N-1) N}{2 (2 (\alpha - N) +1)}
\nn
\\
    \left \langle \chi_{[4]} \right \rangle =
    \frac{a^4 N (N+1) (N+2) (N+3)}{8 (2 (\alpha - N) -3) (2 (\alpha - N) -1)}
& &    \left \langle \chi_{[2,2]} \right \rangle =
    \frac{a^4 (N-1) N^2 (N+1)}{4 (2 (\alpha - N) -1) (2 (\alpha - N) +1)}
\nn
  \ee

  \noindent In general
  \begin{align} \label{eq:character-expansion}
    \boxed{
    \left \langle \chi_{\lambda} \right \rangle
    =
      \frac{\chi_\lambda\{N\}\cdot \chi_\lambda\{\delta_{k,2}\}}{\chi_\lambda\{\delta_{k,1}\}}
      \cdot \frac{a^{|\lambda|}}{P_\lambda(\alpha, N)}
    }
  \end{align}
  where $P_\lambda(\alpha, N)$ is an extra contribution as compared with the short-tailed Hermitian
  Gaussian matrix model case equal to
  \begin{align}
    P_\lambda(\alpha, N) = \prod_{m=1}^{l_\lambda}\prod_{i=1}^{[(\lambda_m+\delta_{m|2})/2]}
    \Big(2(\alpha - N -i+[m/2])+1\Big).
  \end{align}
  \noindent
Here $[...]$ denotes the integer part of a number, and $\delta_{m|2}$ is equal to $1$ for even $m$ and $0$ for odd $m$.
  This product is, in fact, the product over a subset of the Young diagram $\lambda$ boxes
  with coordinates $(i,j)$ that belong to diagonals with odd content $c_{i,j} = i - j$
  of the peculiar combination $2(\alpha - N - c_{i,j}) + 1$.

  There are two, straightforward if a bit tedious, ways to prove
  the superintegrability formula \eqref{eq:character-expansion}.
  One way is via the $\hat W$-representation
  \cite{
    paper:MSh-generation-of-matrix-models-by-w-operators
  }
  and its apparently simple form in the Schur basis
  \cite{paper:MMMR-matrix-model-partition-function-by-a-single-constraint}.
  Another way is via the determinant (Jacobi-Trudi) formula for the Schur polynomials
  and the use of orthogonal polynomials
  \cite{paper:CHPS-orbifolds-and-exact-solutions}.
The third way to prove this formula, probably the easiest one, is to combine the determinant
representations for the matrix model (which is due to its integrable properties, \cite{paper:KMMOZ})
and that for the Schur polynomials as it was done in
\cite[sec.2.2]{paper:MMMR-matrix-model-partition-function-by-a-single-constraint}.
}

Let us note that the matrix integral (\ref{partf0}) lies in the class of Selberg type matrix and eigenvalue models, i.e. those with logarithmic potentials, however, with a specific choice of the integration contours. These models are known \cite{paper:MMS-matrix-model-conjecture-for-exact-BS-periods,Itoyama,paper:MMS-towards-a-proof-of-agt-mm} to possess all basic properties of matrix models but the superintegrability: this later strongly depends on details of the model, and, hence, one had to check it for the concrete matrix Student case.

{\section{Long tails and analytical continuation of factorized formulas} \label{sec:caveats}

After the strongest simplifying property of superintegrability is established,
one can wonder, what  at all is the difference between the long-tail and Gaussian models.
It is, of course, in the physical meaning of exactly-calculable correlators:
they are polynomial in the integration variables, and thus the long-tail correlators
can diverge.
This raises an interesting question of how superintegrability is reflected
in the properties of physically meaningful, convergent correlators,
but here we restrict ourselves to just a brief review of the problem.

  The  superintegrability formula \eqref{eq:character-expansion} in the long-tail case
contains a peculiar factor of $P^{-1}_\lambda$,
which can cause poles in $N$: averages seem to become infinite at certain
  $N = N^*$ and then becomes negative! \
  This is particularly amusing for the quantities that are
  intuitively strictly non-negative, for example
  \begin{align}
    \left \langle \sum_{i=1}^N x_i^2 \right\rangle =
    \frac{a^2 N \left(-2 \alpha  N+2 N^2-1\right)}{(1 + 2 \alpha - 2 N) (-2 \alpha +2 N+1)}
    \mathop{\longrightarrow}_{N \rightarrow \infty} -\frac{a^2 N}{2}
  \end{align}

Recall, however, that the first equality, i.e.
  relation to the Gaussian-distributed matrices
  in Section~\ref{sec:matrix-student-distribution} implies that
  $\alpha = \frac{N + N + n - 1}{2}$, where $N$ and $N + n - 1$ are the dimensions of the matrix $Y$.
  Moreover, the second dimension $M$ should be at least as big as the first one: otherwise, the rank of $Y$ is not sufficient to correctly extract the square root
  and perform the inversion in \eqref{eq:mat-student-density}.
  Putting $\alpha = N + \frac{n - 1}{2}$, one gets
  \begin{align}
    \left \langle \sum_{i=1}^N x_i^2 \right\rangle =
    \frac{a^2 N ((n-1) N + 1)}{n (n - 2)}
    \mathop{\longrightarrow}_{N \rightarrow \infty}  \frac{a^2 (n-1)}{n(n-2)} \cdot N^2
  \end{align}

  \noindent So, there are no longer poles in $N$, and there are no changes as $N$ changes in the sign of the average
  of a strictly non-negative quantity. There is still some denominator, which, however,
  becomes positive and finite starting from some finite $n$; in this example $n > 2$.
  This denominator is the only reminder that the model in question has long tails.

  In other words, positivity is not requires, and is not preserved as soon as one deals with $\alpha$ as a free parameter,
fully independent of $N$.
This once again highlights the importance of finding the right physical degrees of freedom for a model in question.
And, as it often happens with matrix models,
the choice can depend on whether we need physical or mathematical predictions.
The power of matrix model theory should be used with attention and care.
}

\section{Conclusion}

Our conclusion in this paper is that the long-tail distributions can remain
as simple and structured as the exponentially damped ones,
and their non-trivial phase structure can remain well under control.
In particular, various basic properties
from a set of mutually consistent Ward identities (giving rise to the W-representation, \cite{Cassia:2021dpd,paper:MMMR-matrix-model-partition-function-by-a-single-constraint})
to integrability and further superintegrability, do survive, at least in the simplest long-tail example,
that is, the matrix Student distribution.
Remarkably, the long-tail-inspired divergence of integrals for
the correlators, which exhibit integrability in the most straightforward way,
is easily avoided by appropriate analytical continuation
beyond the physical domain, where positivity can be traded for preservation
of the superintegrability.

It remains to understand how general is this result,
and to extend it beyond the Student  distribution analyzed in this paper,
which is long-tail, but still has an apparent Gaussian distribution in the background.
Obvious next steps are to consider $\beta$- and $(q,t)$-deformations,
as well as generalizations to monomial non-Gaussian
and tensor models.

\section*{Acknowledgements}

This work was supported by the Russian Science Foundation (Grant
No.21-12-00400).

\end{document}